\documentclass[AMA,Times1COL]{WileyNJDv5}

\articletype{Research Article}%

\received{Date Month Year}
\journal{Statistics in Medicine}

\usepackage{amssymb}

\usepackage{graphicx}

\usepackage{float}
\usepackage{xcolor}
\usepackage{graphicx}

\usepackage{amsmath}
\usepackage{pdfpages}
\usepackage{tabularx}
\usepackage{multirow}

\DeclareMathOperator{\E}{E}

\newcommand{\N}[2]{\mathcal{N}\left(#1, #2\right)}
\DeclareMathOperator{\I}{\mathcal{I}}

\usepackage{booktabs} 
\usepackage{makecell}

\usepackage{verbatim}

\begin{document}

\title{Predictive Probabilities Made Simple: A Fast and Accurate Method for Clinical Trial Decision Making}

\author[1]{Joe Marion$^*$}

\author[1]{Elizabeth Lorenzi$^*$}

\author[1]{Cora Allen-Savietta$^*$}

\author[1]{Scott Berry}

\author[1,2]{Kert Viele}

\titlemark{Approximate Predictive 
Probabilities}

\authormark{MARION \textsc{et al.}}

\address[1]{\orgname{Berry Consultants}, \orgaddress{\city{Austin}, \state{TX}, \country{USA}}}

\address[2]{\orgname{Department of Biostatistics, University of Kentucky}, \orgaddress{\city{Lexington}, \state{KY}, \country{USA}}}

\corres{*Elizabeth Lorenzi, Berry Consultants, Austin, Texas. \email{elizabeth@berryconsultants.net}}

\abstract[Abstract]{Bayesian predictive probabilities are commonly used for interim monitoring of clinical trials through efficacy and futility stopping rules. Despite their usefulness, calculation of predictive probabilities, particularly in pre-experiment trial simulation, can be a significant challenge. We introduce an approximation for computing predictive probabilities using either a $p$-value or a posterior probability that significantly reduces this burden.
We show the approximation has a high degree of concordance with standard Monte Carlo imputation methods for computing predictive probabilities, and present five simulation studies comparing the approximation to the full predictive probability for a range of primary analysis strategies: dichotomous, time-to-event, and ordinal endpoints, as well as historical borrowing and longitudinal modeling. 
We find that this faster method of predictive probability approximation works well in all five applications, thus significantly reducing the computational burden of trial simulation, allowing more virtual trials to be simulated to achieve greater precision in estimating trial operating characteristics.}

\keywords{Predictive Probabilities, Adaptive Designs, Clinical Trials, Bayesian Designs}

\maketitle

\def\thefootnote{*}\footnotetext{These authors contributed equally to this work}


\section{Background}

Bayesian predictive probabilities are routinely used for interim monitoring of clinical trials, usually as decision quantities for stopping a trial either for anticipated success or futility. Predictive probability based rules may be used with either frequentist or Bayesian primary analyses, and have been shown to work well in comparisons of adaptive trial tools \protect\cite{berry1989monitoring, Broglio2014, Saville2014, Li2023}. Predictive probabilities such as predictive power differ from conditional power in that they incorporate uncertainty about the parameter value \cite{Mehta2011, Pocock1977}.

In some situations predictive probabilities have a simple analytical form (for example the single arm study in Saville et al. \cite{Saville2014}), but in many designs require posterior simulation (see Broglio et al. \cite{Broglio2014}). This posterior simulation creates significant computational overhead when simulating clinical trials. Clinical trial simulation requires simulating thousands of virtual trials, each potentially with multiple interim analysis, each requiring an embedded simulation of the predicted future trial from that interim. This burden can significantly reduce the number of virtual trials that can be simulated, reducing the accuracy of estimated trial operating characteristics.

In contrast, p-values or posterior probabilities may be easily calculated at simulated interim analyses. In this manuscript we propose an approximation of the predictive probability using the current p-value or posterior probability and the current information fraction. In addition to deriving the approximation, we illustrate its use and accuracy in five different common trial designs reflecting a range of endpoints, the use of historical data, and longitudinal modeling. The resulting approximation is accurate and significantly reduces the computational burden, both for the computer and the person who otherwise would need to program the predictive probability.

\section{Derivation of approximate predictive probabilities}

Our approximation is motivated by a frequentist test of a parameter $\theta$. We assume that $\theta$ has been standardized so that the data have unit variance. The null hypothesis $H_0: \theta \leq 0$ will be tested against the alternative $H_a: \theta > 0$ at a particular $\alpha$-level. At the time of the interim analysis there are $n$ patients enrolled, the current information level is $\I_n$, and the primary analysis yields a normally-distributed test statistic $Z_n$ with associated $p$-value $p_n$. The predictive probability $PP_N$ is the probability that the null hypothesis will be rejected if the analysis is performed when $N$ patients have completed the trial. The future information level at that analysis is known to be $\I_N$; the final test statistic $Z_N$ and final $p$-value $p_N$ are unknown when the interim analysis is performed. The predictive probability can be approximated using $p_n$, the $p$-value at the interim, and the information fraction $r = \I_n / \I_N$ assuming $0<\I_n<\I_N$:
\begin{equation}\label{eq:approximation}
        PP\left(p_n, r, \alpha\right) = \Phi\left(\frac{ \Phi^{-1}\left(1-p_n\right) - \Phi^{-1}\left(1-\alpha\right)\sqrt{r}}{\sqrt{1-r}}\right).
\end{equation}

This approximation is derived using standard group-sequential assumptions\cite{Jennison2000, Whitehead1997}. We call this calculation an `approximation' because for most designs the group-sequential assumptions only hold asymptotically and the predictive probability depends upon the choice of prior distribution. Equation~\eqref{eq:approximation} is exact when group-sequential assumptions are met and the prior distribution is $\theta\propto1$. The approximation relies on the test statistic at the interim implying a posterior distribution for the parameter to be tested.
\begin{equation}\label{eq:posterior1}
    \theta \mid \left(Z_n = z_n\right) \sim \N{z_n / \sqrt{\I_n}}{1/\I_n}.
\end{equation}
Under this posterior distribution, the unknown data have the following predictive distribution:
\begin{equation}\label{eq:predictive1}
    Z_{N-n} \mid \left(Z_n = z_n\right) \sim \N{z_n\sqrt{\frac{\I_N-\I_n}{\I_n}}}{\;\;\frac{\I_N}{\I_n}}.
\end{equation}
$Z_{N-n}$ is the test statistic from the unknown data. To compute the predictive probability, we exploit that the final test statistic $Z_N$ can be written as a linear combination of $Z_n$ and $Z_{N-n}$. Then, we use equation~\eqref{eq:posterior1} to compute the probability that $Z_{N-n}$ will be large enough to result in success on the primary analysis test. The complete derivation is given in the Supplemental Appendix. 

\subsection{Approximation for Bayesian analyses}
The approximate predictive probability (aPP) can be applied to Bayesian primary analyses if the posterior distribution is well-approximated by a Gaussian distribution (for example, when the Bernstein-von Mises theorem can be applied). For a Bayesian analysis, the quantity $1-p_n$ in equation~\eqref{eq:approximation} is replaced by the posterior probability of superiority $\mathcal{P}_n$ at the interim and the quantity $1-\alpha$ is replaced by the superiority threshold $\eta$ needed to claim success.
\begin{equation}\label{eq:vayes_approximation}
        PP\left(\mathcal{P}_n, r, \eta\right) = \Phi\left(\frac{ \Phi^{-1}\left(\mathcal{P}_n\right) - \Phi^{-1}\left(\eta\right)\sqrt{r}}{\sqrt{1-r}}\right).
\end{equation}
If a Bayesian analysis includes an informative prior, such as 
historical data or additional information from a longitudinal model, the information fraction may need to be adjusted to reflect the additional information (see Supplemental Appendix). 

\subsection{Applying the approximate predictive probability}
The key assumption underlying the aPP is that the test statistic is (approximately) Gaussian and the information fraction $r$ is known. Common situations where these assumptions hold and the aPP can be applied are given in Table~\ref{tab:examples}.

\begin{table*}[!t]
    \centering
    \caption{Approximating a predictive probability using information level and $p$-levels where $\mathcal{I}_n$ is current information and $\mathcal{I}_N$ is the information at the final analysis.}\label{tab:examples}
    \begin{tabular*}{340pt}{llll}
        \toprule
        Endpoint & Example analyses &  $\mathcal{I}_n$ &  $\mathcal{I}_N$\\
        \toprule
        Continuous & \makecell[lc]{$t$-tests\\ ANOVA/ANCOVA} & Interim sample size & Final sample size\\          
        \cmidrule(lr){1-1}\cmidrule(lr){2-2}\cmidrule(lr){3-3}\cmidrule(lr){4-4}
        Discrete/Categorical & \makecell[lc]{$z$-tests\\ Chi-squared tests} & Interim sample size & Final sample size \\
        \cmidrule(lr){1-1}\cmidrule(lr){2-2}\cmidrule(lr){3-3}\cmidrule(lr){4-4}
        Time-to-event & \makecell[lc]{Log-rank test\\Proportional hazards models} &  Events at interim & Events at final \\            
        \cmidrule(lr){1-1}\cmidrule(lr){2-2}\cmidrule(lr){3-3}\cmidrule(lr){4-4}
        Ordinal/Non-parametric &  \makecell[lc]{Ordinal regression\\ Wilcoxon rank-sum } & Interim sample size & Final sample size  \\
        \cmidrule(lr){1-1}\cmidrule(lr){2-2}\cmidrule(lr){3-3}\cmidrule(lr){4-4}
        Count data &  \makecell[lc]{Generalized linear regressions\\(e.g., Poisson regression)} & Interim exposure &  Final exposure\\
       \bottomrule
    \end{tabular*}
\end{table*}

The normality assumption is commonly satisfied and includes the class of maximum likelihood estimators under sufficient regularity conditions\cite{Vaart1998}. For many analyses, the normality of the test statistic relies on key assumptions about the underlying data. For instance, the two-proportion $z$-test is only approximately Gaussian when there is sufficient size to ensure an adequate number of successes and failures. When the sample size is inadequate or when success/failure probabilities approach extremes (0 or 1), both the $z$-test and predictive probabilities computed using that test may be inappropriate (see Supplemental Appendix). 

The aPP also assumes that the covariance between the interim and final test statistics is known and given by the square root of information ratio $\mathcal{I}_n / \mathcal{I}_N$. For many analyses, the information is simply the number of complete observations or sample size. However, the design of the study, including definitions of primary analysis population, and handling of missing data and dropouts should be considered when determining the information. For example, a study using a modified intent-to-treat primary analysis may have a final sample size that is smaller than the planned maximum.

\subsection{Estimating the information}

For some designs, the aPP
may be difficult to apply because either $\mathcal{I}_n$ or $\mathcal{I}_N$ are unknown. For example, the maximum number of events may not be known in a time-to-event 
trial with a fixed follow-up duration. An interim analysis using a repeated measures model could include early visit data for patients with incomplete follow-up, but how should they be counted toward the information? Bayesian methods, such as those that include informative priors, external information, or additional endpoints, may also contribute information in ways that are difficult to quantify.  

To use the aPP
in these settings, it may be necessary to estimate $\mathcal{I}_n$, $\mathcal{I}_N$, or both. For a time-to-event 
analysis with an unknown number of final events, the interim data could be used to estimate the number of events when the trial is complete. If a simple solution is unavailable, we propose a generic approach that can be widely applied using the effective sample size.
This approach estimates the unknown information by comparing to a similar, reference analysis where the information is known\cite{Satoshi2008, Neuenschwander2010, Viele2014}. Let $\hat{\theta}$ be the treatment effect estimated by the analysis with unknown information, $\tilde{\theta}$ be an estimate of the treatment effect from the reference analysis, and $\tilde{\mathcal{I}}$ be the information of the reference analysis. Assuming that the variance of both estimates is proportional to the information level, the unknown information can be estimated using the variance ratio:
\begin{equation}\label{eq:corrected}
    \hat{\mathcal{I}} = \frac{\text{Var}( \tilde{\theta})}{\text{Var}( \hat{\theta})} \cdot \tilde{\mathcal{I}}.
\end{equation}
%
The success of this approach depends on the suitability of the reference analysis, which should be comparable to the analysis of interest, and the assumption that the variance is proportional to the information level. 
We include two examples where we estimate the information level using 
the effective sample size: an analysis with a longitudinal model informed by an early endpoint 
and an analysis that dynamically borrows information from an external control.

\subsection{Interpretation of the approximation}

\begin{figure*}[t]
\centering
\includegraphics[width=500pt]{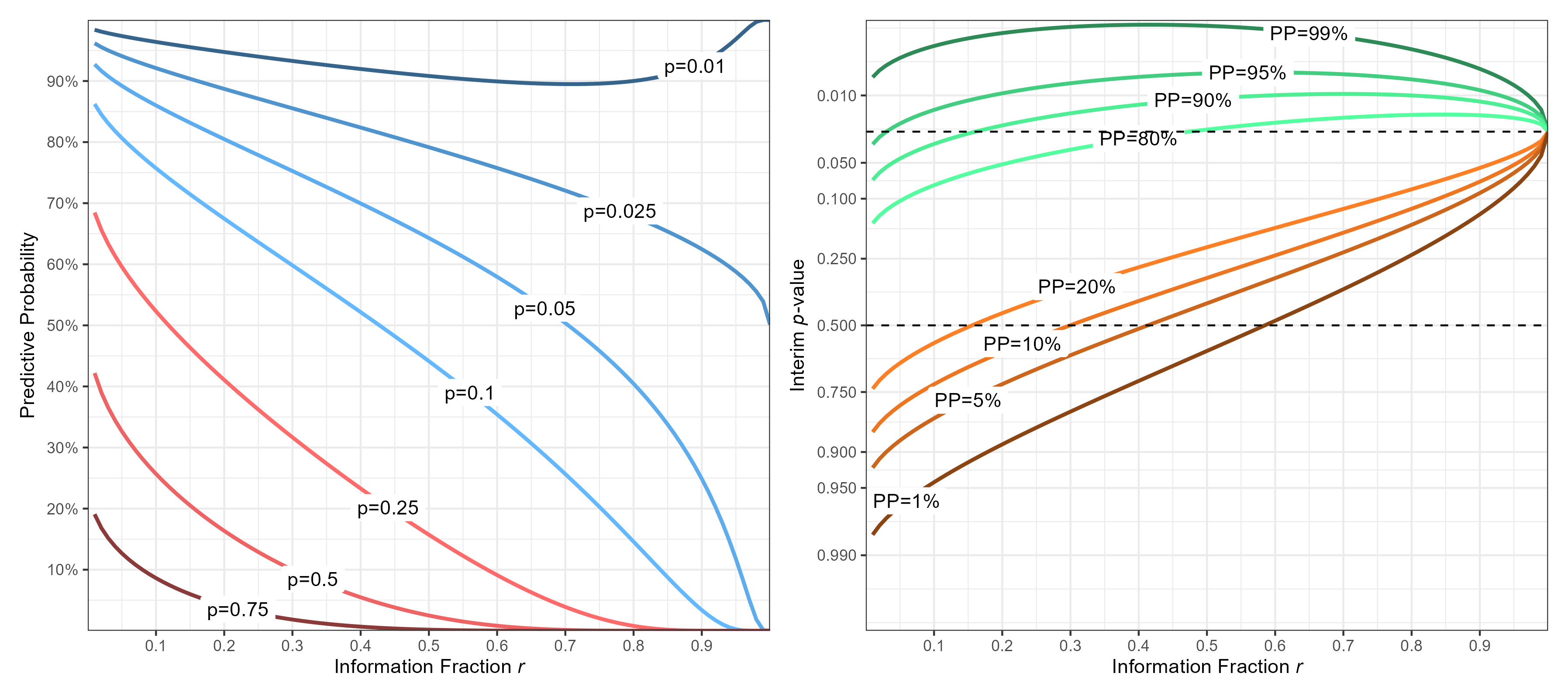}
\caption{Relationship between approximate predictive probability and $p$-values as a function of the information fraction $r$ with $\alpha=0.025$. The left panel shows the predictive probability for several $p$-values; the right panel shows $p$-values for several predictive probabilities, obtained by inverting equation~\eqref{eq:approximation}. The y-axis of the right panel is shown on the probit scale; the dashed lines show key values $p=0.50$ and $p=0.025$.}\label{fig:interpretation}
\end{figure*}

The aPP provides insight into the relationship between $p$-values and predictive probabilities. Figure~\ref{fig:interpretation} shows equation~\eqref{eq:approximation} for fixed values of $p_n$ (left panel) and $PP$ (right panel) plotted as a function of $r$ with $\alpha=0.025$. The left panel shows that when the $p$-value is greater than the success threshold $\alpha$, predictive probabilities decrease as the information fraction increases. This makes sense intuitively as with less outstanding information there is less opportunity for the data to improve sufficiently to meet the success criteria. The exception to this rule is a case in which the $p$-value is less than the target threshold ($p=0.01$ in the figure). In this instance, the predictive probability starts to increase as the information fraction approaches one. This reflects the reduced possibility that the interim data may worsen due to sampling variability when the primary analysis is performed. Paradoxically, this means that when the interim data are strongly positive, the predictive probability of success at the current $N$ may exceed the predictive probability of success at $N_{max}$.

The right panel of Figure~\ref{fig:interpretation} presents the inverse of equation~\eqref{eq:approximation}, showing the $p$-values that are equivalent to predictive probabilities and could trigger interim decisions at each information fraction. The green lines show potential early success rules. For these rules, the $p$-value needed to trigger stopping is often smaller than the final $\alpha$-level, accounting for the possibility that the strength of the data may decrease as outstanding information accumulates. For designs with type I error control, the $p$-values needed to stop enrollment for success are often smaller than their group-sequential equivalents, as interim analyses are not the definitive analysis: predictive probability designs require success when all the data become available.

The orange lines in the right panel show common futility-stopping boundaries. As the information fraction increases, the $p$-value needed to avoid futility approaches the target $\alpha$. The line for $p=0.50$ shows the predictive probability of success when the data are balanced between evidence of benefit and harm. This provides a measure of each futility rule's aggressiveness. Futility is triggered when $r>0.15$ for a 20\% threshold, when $r>0.29$ for a 10\% threshold, and when $r>0.41$ for a 5\% threshold. The $1\%$ rule is particularly conservative: when the treatment and control arms are equivalent, more than half the information should be complete to trigger stopping for futility.

\section{Example: the SHINE trial}
We demonstrate our aPP tool using the SHINE clinical trial\cite{SHINE}, which compared intensive glucose control to the standard of care for stroke patients with hyperglycemia. The primary endpoint was a modified Rankin score measured at 90 days, dichotomized into favorable and non-favorable outcomes. While the trial was implemented as a group-sequential design, the study was re-analyzed using a Goldilocks design as the ``Shadow SHINE" trial\cite{Broglio2022}. The alternative design used interim analyses beginning at 500 patients enrolled and at every 100 patients thereafter until the maximum sample size of 1400 patients was enrolled.  At each interim analysis, two predictive probabilities were calculated: 1) $PP_N$, the predictive probability at the current sample size $N$; and 2) $PP_{max}$, the predictive probability at the maximum sample size. Enrollment would stop for anticipated success at sample size $N$ if $PP_N > 0.99$ or for futility if $PP_{max} < 0.05$. The results of the alternative interim analyses are shown in Table~\ref{tab:shadow_shine}; estimates of $PP_N$ have been omitted because they are below 0.01 for each interim and predictive probability method.

\begin{table*}[!h]
    \centering
    \caption{Predictive probabilities for the Shadow SHINE trial.  The number of patients enrolled and number of patients with 90-day outcomes is given by $N$ and $n$, respectively.  Observed response rate $p$-values at the interim are given in the columns labeled Treatment, Control, and $p_n$.  The last two columns show the predictive probabilities for Shadow SHINE interim analyses.} \label{tab:shadow_shine}
    \begin{tabular*}{223pt}{ccccccccccc}
        \toprule
        \multicolumn{5}{c}{Interim Data} & \multicolumn{2}{c}{$PP_{max}$} \\
        \cmidrule(lr){1-5}\cmidrule(lr){6-7}
         $N$ & $n$ & Treatment & Control & $p_n$ & iPP & aPP \\
         \cmidrule(lr){1-5}\cmidrule(lr){6-7}
        498 & 432 & 0.264 & 0.244 & 0.3535 & 0.194 & 0.182 \\
        579 & 515 & 0.256 & 0.221 & 0.2058 & 0.372 & 0.349 \\
        700 & 621 & 0.250 & 0.232 & 0.3372 & 0.125 & 0.129 \\
        800 & 715 & 0.231 & 0.228 & 0.4994 & 0.028 & 0.026 \\
        \bottomrule
    \end{tabular*}
\end{table*}
For the Shadow SHINE re-analysis, the results of the iPP and aPP analysis are similar and would lead to the same trial decisions. Note that the iPP results differ slightly from the original\cite{Broglio2022} because we have omitted the longitudinal model from the iPP to simplify the comparison.

\section{Simulation studies}

In simulation studies we explore the approximation's performance against a Bayesian model-imputed predictive probability. Our goal is to determine
\begin{itemize}
\item the similarity between aPP and iPP; and
\item whether the aPP and iPP lead to similar decisions within a clinical trial design even when they do not match exactly.
\end{itemize}

We simulate individual-level data for three endpoint types (dichotomous, time-to-event, and ordinal) and study designs. Then, we present an extension of the ordinal example using more complex analyses: historical borrowing and longitudinal modeling primary analyses.

\subsection{Dichotomous and Time To Event Primary Endpoints}

Simulated trials with dichotomous and time-to-event endpoints have a maximum sample size of 500 and interim Goldilocks-style sample size re-estimations at 300 and 400 participants enrolled \cite{Broglio2014}. 

At each interim sample size re-estimation, design decisions are made based on the aPP and Bayesian iPP. The prespecified study design makes one of three decisions:
\begin{itemize}
\item Stop trial
enrollment for expected success if $PP_N > 90\%$);
\item Stop trial enrollment 
for futility if $PP_{max} < 5\%)$; or
\item Continue trial enrollment.
\end{itemize}

For the time-to-event scenario, we assume the maximum follow-up per subject is one year. The final analysis is not timed based on the number of events but rather the length of follow-up, so we approximate the final number of events for $I_N$ by calculating the expected number of events at full follow-up using the estimated hazard rate at the interim. 

We follow the procedure described in Table~\ref{tab:examples} for calculating the aPP. For the traditional Bayesian imputed method, we follow the procedure described by Broglio et al.\cite{Broglio2014}. Additional details on accrual and information fraction can be found in the supplement.

\subsection{Ordinal primary endpoint}\label{sec:ordinal}
We simulate a two-arm, 1:1 randomized controlled trial with a 6-level ordinal primary outcome. The trial has a maximum sample size of $N=1500$ with interims when $N=500, 750, 1000$ and $1250$ subjects have enrolled. At each interim, the design can make one of three decisions: stop trial enrollment for expected success if $PP_N > 90\%$; stop trial enrollment for futility if $PP_{max} < 10\%$; or continue trial enrollment. Additional details on the simulation assumptions can be found in the supplement.

\subsection{Dichotomous, Time To Event, and Ordinal Results}\label{sec:endpointresults}

\begin{figure*}[!htbp]
\centering
\includegraphics[width=500pt]{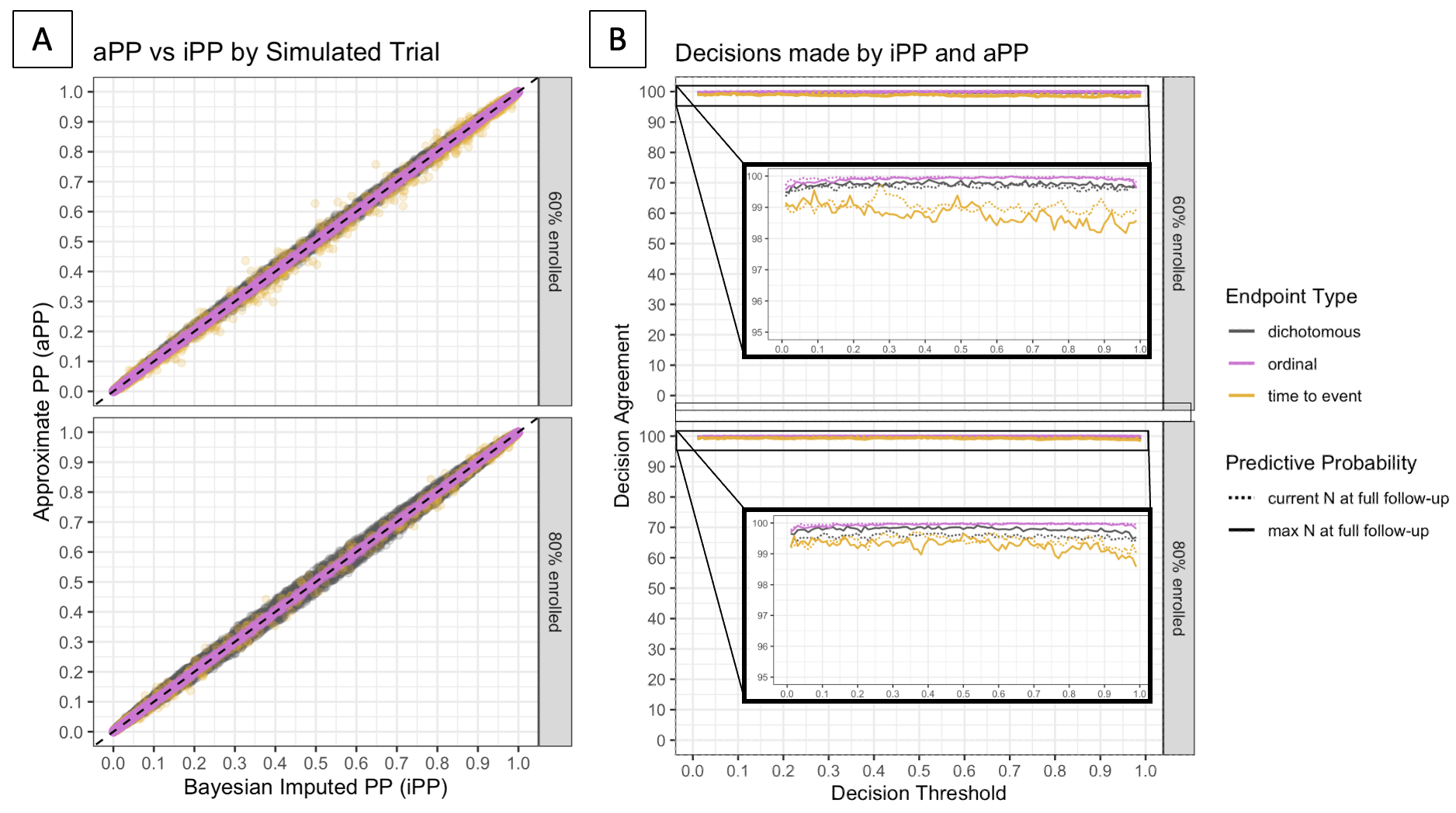}
\caption{Comparison of imputed predictive probability (iPP) and approximate predictive probability (aPP) comparing the dichotomous, time-to-event, and ordinal endpoint examples. Panel A summarizes the aPP compared to the iPP for the interim timed when 60\% are enrolled (top) and the interim timed when 80\% are enrolled (bottom). Panel B shows the proportion of simulations where the iPP and aPP make the same decision based on the threshold along the x-axis.}\label{fig:pp_combined}
\end{figure*}

Figure \ref{fig:pp_combined} displays a comparison of the aPP and iPP in panel A for each of the three examples for interim analyses timed when approximately 60\% and 80\% of patients have been enrolled. In the ordinal example, we include the interim timed when 1000 patients have been enrolled and 1250 have been enrolled out of a maximum of 1500 (resulting in 66.6\% and 83\% enrolled). Points along the diagonal line represent simulations where the aPP and iPP are the same. Across the three examples, almost all points lie along the diagonal line, with a few time-to-event example simulations deviating by at most 0.05 in either direction at the 60\% enrolled interim analysis. 

In panel B, we show the proportion of simulations where the aPP and iPP (y-axis) make the same decision for a range of decision thresholds (x-axis). At the first interim when approximately 60\% have been enrolled, the aPP and iPP decisions agree on over 98\% of simulations regardless of the threshold. Decisions for the ordinal and dichotomous cases agree over 99\% of the time. At the interim when approximately 80\% of patients have been enrolled, the iPP and aPP make the same decision over 98.5\% of the time for the three examples. The time-to-event example has a few cases where the aPP and iPP are slightly different, resulting in a slightly lower proportion of simulations where decisions agree. This is likely due to the estimation of the information going into the approximation. Specifically, more deviations occur at the earlier interim because the estimate of the hazard rate may be noisier with less data. In general, the aPP is very similar to the iPP across the three endpoint types. The high degree of agreement between the two approaches indicates that for these designs the operating characteristics using aPPs would be nearly identical to those using iPPs.

\subsection{Ordinal primary endpoint modeling extensions}\label{sec:ordinal_extensions}

We provide two variations that use different modeling approaches. The first is an analysis 
that uses 30-day outcomes to predict 90-day outcomes through 
longitudinal modeling; the second is an analysis model that incorporates an external control cohort through Bayesian dynamic borrowing. 

\subsubsection{Longitudinal modeling}\label{sec:longitudinal}

We use Bayesian ordinal logistic regression to jointly model 90-day and 30-day outcomes. When the early measurement is predictive of the final measurement, longitudinal modeling can improve the precision of the treatment estimate and improve clinical trial decision making. Predictive probabilities are computed by imputing datasets with complete 90-day outcomes using the longitudinal model and then applying the primary analysis to each imputed dataset. This procedure is computationally expensive due to the cost of repeatedly fitting the primary analysis.

We can apply the aPP using the posterior probability of superiority from the longitudinal model, the number of complete 90-day outcomes $n$, and the future sample size (either $N$ or $N_{max}$). However, the incorporation of 30-day outcomes may increase the amount of information beyond the nominal level $n$. To address this concern, we estimate the effective sample size shown in equation~\eqref{eq:corrected}. First, we perform the analysis incorporating the longitudinal model, providing an estimate $\hat{\theta}$ of the treatment effect. Then, we repeat the analysis without the longitudinal component, yielding the estimate $\tilde{\theta}$.  

\subsubsection{Borrowing external data}\label{sec:borrowing}
We assume there is a historical cohort 
of 500 total subjects, with 250 randomized to the treatment arm and 250 randomized to control. The primary analysis model is a Bayesian cumulative logistic regression model that estimates an odds ratio of treatment relative to control with a hierarchical prior that borrows information from the historical cohort's treatment effect. Within the hierarchical prior, we estimate the variance parameter of the two treatment effects to allow their similarity to drive the degree of borrowing (dynamic borrowing \cite{Viele2014}). See Supplementary Appendix for the full model specification.

The calculation of imputed predictive probabilities in this setting requires imputing outcomes using the estimates from the Bayesian borrowing model and then applying the same primary analysis model to each imputed dataset. This requires fitting a model with MCMC thousands of times based on the number of posterior samples.

The aPP 
provides a huge computational advantage in this setting. It requires a single run of the model where we output a posterior probability of benefit (probability the odds ratio
$>$ 1), the current amount of information, and the future sample size (either $N$ or $N_{max}$). However, summarizing the current amount of information and the final amount of information is not straightforward. 
With dynamic borrowing, the amount of information is not fixed but depends on the variance parameter estimated of the two treatment effects. As the variance of the effects gets smaller, the model borrows more information from the external cohort. 

We estimate the effective sample size described in equation \ref{eq:corrected} to summarize the information gained from dynamic borrowing by estimating the treatment effect without the use of dynamic borrowing under a neutral prior to obtain $\hat{\theta}$ and again under the proposed model with the dynamic borrowing prior to yield the estimate $\tilde{\theta}$. Using these estimates, we calculate $\hat{n}$. We can measure the number of observations borrowed by taking the difference $n_{bor} = \hat{n} - n$. This difference carries through to the expected total sample size $N$ or $N_{max}$ under the assumption that the analysis will borrow a similar amount of information at the final analysis at $N$. 

\subsubsection{Ordinal primary endpoint modeling extension results}
We compare three approaches to computing predictive probabilities for the longitudinal modeling and borrowing designs: 1) the 
iPP, 2) the aPP
using the nominal information fraction (nPP), and 3) the aPP 
using the estimated information fraction (ePP). For the longitudinal modeling example, the nPP
summarizes the number of observations with complete follow-up at the interim and ignores the observations with incomplete follow-up. For the Bayesian borrowing example, the nominal information fraction ignores the data used from the external cohort. The goals of these simulations are to show whether the aPP
results in similar predictive probabilities to the imputation-based approach and whether estimating the information fraction improves the accuracy of the predictive probabilities.

\begin{figure*}[!htbp]
\centering
\includegraphics[width=.8\textwidth]{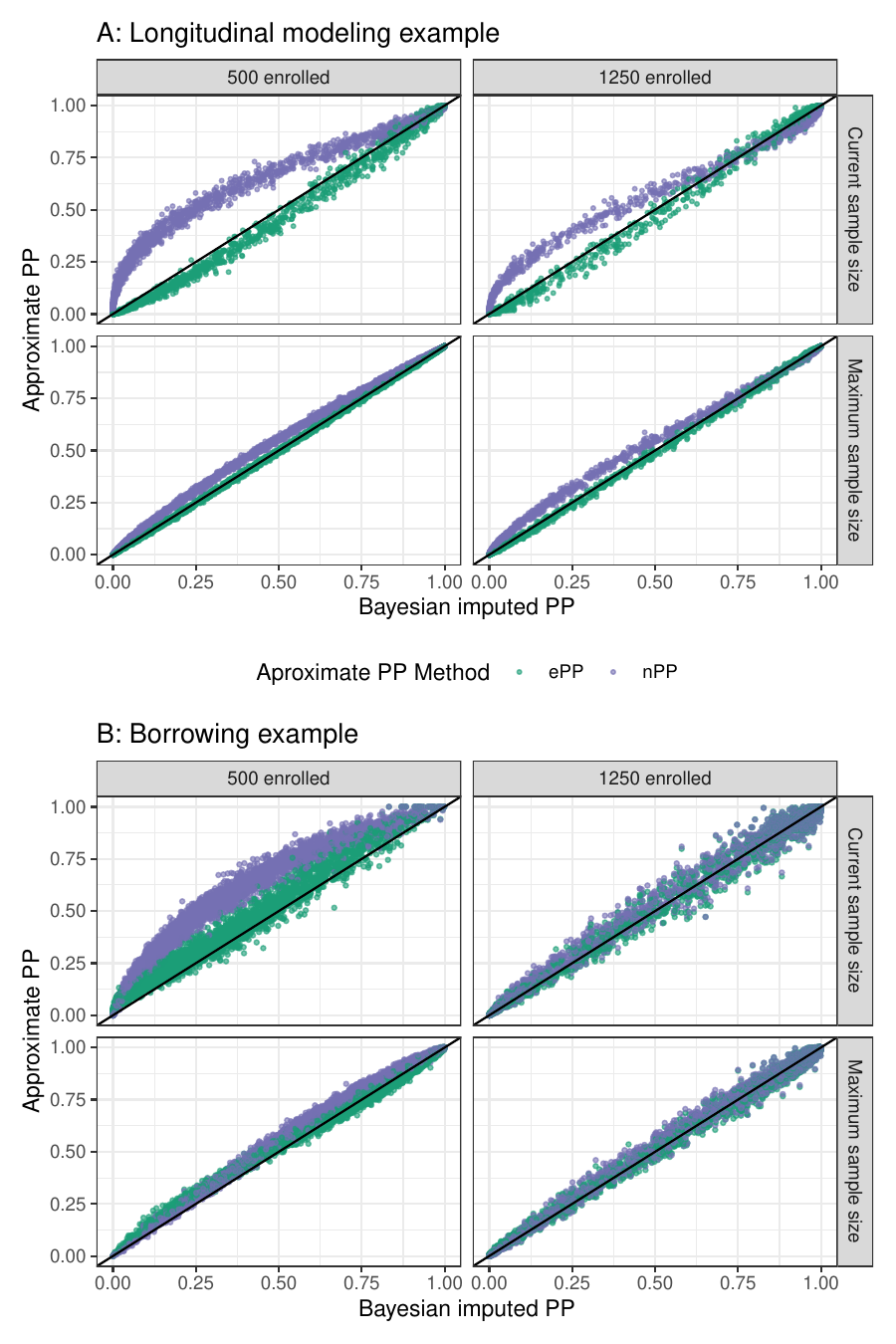}
\caption{Comparison of imputed predictive probability (iPP) and approximate predictive probability (aPP) for the longitudinal model (A) and borrowing model (B) examples. Two versions of the approximation are computed, one using the nominal information fraction (nPP) and the other using the estimated information fraction (ePP).}
\label{fig:pp_ordinal_bor_long}
\end{figure*}

Figure \ref{fig:pp_ordinal_bor_long} displays the estimates of the predictive probabilities for the current sample size (top row of graphs) and the maximum sample size (bottom row of graphs) estimated at the interims when 500 and 1250 subjects have been enrolled (columns) for the longitudinal example in panel A and the borrowing example in panel B. The approximate and imputed predictive probabilities differ most when the number of enrolled patients with incomplete follow-up is large relative to the maximum information. This difference is maximized for $PP_N$ at the first sample size. The 
nPP (purple dots) is likely to overestimate the predictive probability due to underestimating the current information. This problem is less pronounced for predictive probabilities near the decision boundaries, so the overall impact on the operating characteristics of the design is limited. The 
ePP has better concordance with the iPP, though in the longitudinal example it tends to slightly overestimate the interim information, leading to slightly lower predictive probabilities. In the borrowing example, the ePP slightly overestimates the predictive probability at early interims due to underestimating the current information. 


Table~\ref{tab:longitudinal_results} shows the probability of stopping at each interim analysis for all three predictive probability methods for the two modeling examples. The operating characteristics using iPP and ePP are nearly identical. The operating characteristics using nPP show some differences, primarily at the first interim analysis where underestimating the information fraction leads to less stopping for futility and more for success. 

\begin{table*}[!h]
    \centering
    \caption{Comparing interim decisions for the longitudinal and borrowing ordinal examples. For each interim, we summarize the proportion of trials that make a decision (success or futility) based on the imputed predictive probability (iPP), the approximate predictive probability using the estimated information fraction (ePP), and the approximate predictive probability using the nominal information fraction (nPP) at two odds ratios (OR).}\label{tab:longitudinal_results}
    \begin{tabular*}{400.5pt}{cccccccccccccccccccc}
    \toprule
    \multicolumn{2}{c}{ } & \multicolumn{3}{c}{OR=1.0} & \multicolumn{3}{c}{OR=1.2} & \multicolumn{3}{c}{OR=1.4} & \multicolumn{3}{c}{OR=1.6} \\
    \cmidrule(l{3pt}r{3pt}){3-5} \cmidrule(l{3pt}r{3pt}){6-8}  \cmidrule(l{3pt}r{3pt}){9-11} \cmidrule(l{3pt}r{3pt}){12-14}
    Model & Interim & iPP & ePP & nPP & iPP & ePP & nPP & iPP & ePP & nPP & iPP & ePP & nPP\\
    \cmidrule(l{3pt}r{3pt}){1-1} \cmidrule(l{3pt}r{3pt}){2-2}  \cmidrule(l{3pt}r{3pt}){3-5} \cmidrule(l{3pt}r{3pt}){6-8}  \cmidrule(l{3pt}r{3pt}){9-11} \cmidrule(l{3pt}r{3pt}){12-14}
    \multirow{5}{*}{Longitudinal}  & 1 & 0.50 & 0.50 & 0.43 & 0.15 & 0.15 & 0.12 & 0.10 & 0.11 & 0.11 & 0.25 & 0.24 & 0.29\\
     & 2 & 0.20 & 0.20 & 0.20 & 0.13 & 0.13 & 0.12 & 0.23 & 0.23 & 0.21 & 0.41 & 0.41 & 0.37\\
     & 3 & 0.13 & 0.13 & 0.14 & 0.15 & 0.14 & 0.13 & 0.28 & 0.28 & 0.26 & 0.24 & 0.24 & 0.23\\
     & 4 & 0.09 & 0.09 & 0.10 & 0.17 & 0.17 & 0.15 & 0.21 & 0.21 & 0.20 & 0.09 & 0.09 & 0.09\\
     \cmidrule(l{3pt}r{3pt}){3-5} \cmidrule(l{3pt}r{3pt}){6-8} \cmidrule(l{3pt}r{3pt}){9-11} \cmidrule(l{3pt}r{3pt}){12-14}
     &Total & 0.93 & 0.92 & 0.88 & 0.61 & 0.59 & 0.52 & 0.83 & 0.82 & 0.78 & 0.98 & 0.98 & 0.97\\
    \hline \hline
    \multirow{5}{*}{Borrowing} & 1 & 0.07 & 0.05 & 0.07 & 0.02 & 0.02 & 0.04 & 0.02 & 0.04 & 0.06 & 0.05 & 0.07 & 0.12\\
     & 2 & 0.22 & 0.20 & 0.21 & 0.09 & 0.08 & 0.09 & 0.17 & 0.16 & 0.18 & 0.40 & 0.38 & 0.39 \\
     & 3 & 0.22 & 0.23 & 0.22 & 0.11 & 0.13 & 0.12 & 0.29 & 0.28 & 0.26 & 0.34 & 0.34 & 0.30 \\
     & 4 & 0.20 & 0.22 & 0.20 & 0.18 & 0.19 & 0.18 & 0.27 & 0.27 & 0.25 & 0.16 & 0.16 & 0.15 \\
      \cmidrule(l{3pt}r{3pt}){3-5} \cmidrule(l{3pt}r{3pt}){6-8}  \cmidrule(l{3pt}r{3pt}){9-11} \cmidrule(l{3pt}r{3pt}){12-14}
    & Total & 0.71 & 0.69 & 0.70 & 0.41 & 0.42 & 0.43 & 0.74 & 0.75 & 0.75 & 0.95 & 0.95 & 0.96 \\
    \hline
    \bottomrule
    \end{tabular*}
\end{table*}

\section{Discussion}

The aPP is a user-friendly alternative to more computationally-intensive imputed predictive probabilities. The aPP can be confidently employed in the common encountered circumstance where the test statistic or posterior distribution is approximately Gaussian and the information fraction is known or well-estimated. The computational savings are large. In the Bayesian borrowing example described above, the iPP calculation took 15 minutes (single-threaded on a 2.2GHz Intel Core i7 processor with 1000 MCMC samples) compared to near-instantaneous calculation using the aPP approach. Additionally, the method is simple to implement, without any intricate coding imputing potential future data.

We have shown a range of examples in which aPP and iPP have good, but not perfect, agreement. When the information fraction is not explicit, the aPP does not exactly align with the iPP; however, estimating the information fraction can allow the aPP to be applied in these settings (time-to-event, borrowing, and longitudinal modeling cases). In cases where the nominal information may be incorrect, such as the longitudinal and borrowing examples, adjusting the information fraction using the effective sample size improves the performance of the aPP. Still, even after these adjustments, small discrepancies between the aPP and the imputed gold standard persist. 

While good agreement with iPP is ideal, it is not a prerequisite for utilizing the aPP. Regardless of the type of predictive probability used, a clinical trial design should be justified by its simulated operating characteristics. The computational methods used during the conduct of a clinical trial should be consistent with those used in the design phase. Like any approximation, if the approach is prespecified and the simulations accurately reflect the use of aPP, the resulting design is valid.

There remain endpoints/analyses where the approximation may be particularly difficult to apply. Consider the Finkelstein-Schoenfeld test\cite{Finkelstein1999}, which has been used in clinical trials for amyotrophic lateral sclerosis\cite{Berry2013} and cardiovascular outcomes\cite{Ferreira2020}. Depending on the construction of the endpoint, comparisons between individual patients may change over time, meaning that the current information is unstable and likely to change over time. This makes it difficult to apply the aPP because the correlation between the interim and final test statistics is unclear. Currently, the best way of evaluating this correlation is through simulation and modeling, meaning that imputation may be the preferred method of computing predictive probabilities for this composite win-ratio endpoint. 

The connection between predictive probabilities, $p-$values and information fraction opens new possibilities for developing Goldilocks designs. Instead of simulation based methods, group-sequential calculations could be used to evaluate the operating characteristics of a Goldilocks design. Consequently, there's potential for leveraging group-sequential theory to devise Goldilocks designs that have analytic type I error control.  

Thus, while not a panacea, the approximation should allow greater use of predictive probabilities in practical clinical trial design.



\bmsection*{Acknowledgments}
The authors would like to thank Val Durkalski, Will Meuer, and Kristine Broglio for providing additional details about the SHINE trial.  They also thank their colleagues at Berry Consultants for many insightful discussions.


\bmsection*{Conflict of interest}
The authors are employees of Berry Consultants, a consulting company specializing in the design of complex Bayesian adaptive clinical trials.

\pagebreak

{\Large \textbf{Supplement to \textit{Predictive Probabilities Made Simple: A Fast and Accurate Method for Clinical Trial Decision Making}}}

\section{Derivation of the Approximation}\label{sec:derivation}
Following~\cite{Jennison2000, Whitehead1997} we assume that the test-statistics $(Z_n, Z_N)$ follow the canonical joint distribution with the following properties.

\begin{enumerate}
    \item $\left(Z_n, Z_N\right)$ are bivariate normal.
    \item $\E\left(Z_n\right) = \theta \sqrt{\I_n}$ and $\E\left(Z_N\right) = \theta \sqrt{\I_N}$.
    \item $\text{Var}\left(Z_n\right) = \text{Var}\left(Z_N\right) = 1 $ and $\text{Cov}\left(Z_n , Z_N\right) = \sqrt{\I_n / \I_N}$.
\end{enumerate}
Under these assumptions we can decompose the final test statistic as the weighted sum of $Z_n$ and another independent Gaussian random variable $Z_{N-n}$ which comes from the unknown data.
\begin{equation}\label{eq:decompose}
    Z_N \sqrt{I_N} = Z_n \sqrt{\I_n} + Z_{N-n} \sqrt{\I_N - \I_n}
\end{equation}
\begin{equation}\label{eq:sampling}
    Z_{N-n} \sim \N{\theta\sqrt{\I_N - \I_n}}{1}
\end{equation}
To obtain the predictive distribution of $Z_N$, we assume an uninformative prior distribution $\theta\propto 1$ which yields the posterior distribution
\begin{equation}\label{eq:posterior}
    \theta \mid \left(Z_n = z_n\right) \sim \N{z_n / \sqrt{\I_n}}{1/\I_n}.
\end{equation}
The predictive distribution for $Z_{N-n}$ is obtained by combining equations~\eqref{eq:sampling} and~\eqref{eq:posterior}
\begin{equation}\label{eq:predictive}
    Z_{N-n} \mid \left(Z_n = z_n\right) \sim \N{z_n\sqrt{\frac{\I_N-\I_n}{\I_n}}}{\;\;\frac{\I_N}{\I_n}}.
\end{equation}
When all the data are complete, the hypothesis test will be performed at the one-sided $\alpha$-level and success is declared if $Z_N > \Phi^{-1}\left(1-\alpha\right)$.  Re-arranging~\eqref{eq:decompose} gives the success criteria conditional on the available data $Z_n=z_n$.
\begin{equation*}
    \begin{split}
        \frac{ z_n \sqrt{\I_n}  + Z_{N-n} \sqrt{\I_N - \I_n}}{\sqrt{\I_N}}  &> \Phi^{-1}\left(1-\alpha\right) \\
        Z_{N-n}  \sqrt{\I_N - \I_n} &> \Phi^{-1}\left(1-\alpha\right)\sqrt{\I_N} -  z_n \sqrt{\I_n} \\
        Z_{N-n}   &> \Phi^{-1}\left(1-\alpha\right)\sqrt{\frac{\I_N}{\I_N - \I_n}} -  z_n \sqrt{\frac{\I_n}{\I_N - \I_n}} \\
    \end{split}
\end{equation*}
The predictive probability of exceeding this threshold follows from the predictive distribution~\eqref{eq:predictive}.
\begin{equation*}
    \begin{split}
        PP &= \Pr \left(  Z_{N-n}   > \Phi^{-1}\left(1-\alpha\right)\sqrt{\frac{\I_N}{\I_N - \I_n}} -  z_n \sqrt{\frac{\I_n}{\I_N - \I_n}} \;\Biggr\lvert\; Z_n = z_n\right) \\
        &= \Phi\left( \frac{ z_n \sqrt{\frac{\I_n}{\I_N - \I_n}} + z_n\sqrt{\frac{\I_N-\I_n}{\I_n}} - \Phi^{-1}\left(1-\alpha\right)\sqrt{\frac{\I_N}{\I_N - \I_n}}}{\sqrt{\I_N/\I_n}} \right)\\
        &= \Phi\left( z_n \sqrt{\frac{\I_N}{\I_N-\I_n}} -\Phi^{-1}\left(1-\alpha\right)\sqrt{\frac{\I_n}{\I_N - \I_n}}      \right).\\
    \end{split}
\end{equation*}
Letting $r = \I_n/\I_N$ and substituting $z_n = \Phi^{-1}\left(1-p\right)$ gives the main result. 

\section{Simulation Study Methodology Details}
We provide additional methodological details for the simulation studies presented in the main paper. 

In both the dichotomous and time-to-event simulation studies, we generated 10,000 simulated trials. These simulations were run in FACTS Version 6.5 (Berry Consultants, Austin, Texas, 2023) and processed using R Statistical Software (v4.3.0; \cite{R}). 

\subsection{Dichotomous Primary Endpoint}
\label{sec:dichotomous}

The primary endpoint is dichotomous, where 1 indicates a negative outcome, measured at three months. The primary analysis is a test of two proportions. 

In all scenarios, we assume an annual rate of 0.50 for the control arm, then explore two possible two treatment effects: 0 and 0.15 reduction from this annual rate. This leads to scenarios of 0.50 in both groups or 0.50 and 0.35 in the control and treatment groups, respectively.

The study has a maximum sample size of 500. There are interim Goldilocks-style sample size re-estimations at 300 and 400 participants enrolled. 
We assume an accrual rate of 5 patients per week. This results in approximately 235 patients who have observed their 3-month endpoint at the first interim and 335 at the second interim. 

\subsection{Time To Event Details}\label{sec:tte}

The setup mimics a two-arm randomized trial in oncology where patients are equally randomized to a control or treatment arm. The primary endpoint assumes an event is a negative outcome measured with a maximum of one year of follow-up (e.g., progression-free survival). The primary analysis is a log-rank test. 

The maximum follow-up per subject is one year. We assume the one-year event rate on control is 30\%. 
Four treatment effect scenarios are considered: when the hazard ratio is 0.6, 0.7, 0.8, and 1.0 (null). 

We assume an accrual rate of 5 patients per week. 
Under this accrual, approximately 40 patients will have had the opportunity to complete one year of follow-up at the first interim analysis, and approximately 140 patients at the second interim analysis.


As the final analysis is not timed based on the number of events but rather the length of follow-up, we have to approximate the final number of events for $I_N$ by calculating the expected number of events at full follow-up using the estimated hazard rate at the interim.

\subsection{Ordinal simulation details}
Our ordinal example is a two-arm, 1:1 randomized controlled trial with a 6-level ordinal primary outcome measured at 90 days. A cumulative logistic regression model estimating a proportional odds ratio is used. The trial has a maximum sample size of $N=1500$ with interims when $N=500, 750, 1000$ and $1250$ subjects have enrolled. Simulations are performed for odds ratios of 1.6, 1.4, 1.2 and 1.0 (null).  
Accrual is simulated at a rate of two patients per day. Under this accrual, at each interim analysis there are an average of 180 enrolled patients who have not completed the 90-day follow-up, 120 of whom have 30-day measurements. 
For each scenario, 1000 simulations are performed. They were run and processed using R Statistical Software (v4.3.0; \cite{R}).

\section{Dynamic Borrowing Model}\label{sec:app_bor_model}

In the ordinal example in borrowing section of the main paper,  we use the following Bayesian primary analysis model that incorporates the ongoing trial data with the historical control cohort. Let $Y_i \in \{0, 1,2,3,4,5\}$ denote the ordinal endpoint for patient $i = 1, ...., N$, where $N$ is the total number of subjects in the current trial and external cohort. Let $\gamma_{ij}$ denote the probability of patient $i$ observing an mRS category of $j$ or lower, $\gamma_{ij} = Pr(Y_i \le j)$. For $j = 0, 1, 2, 3, 4$, the model specification is as follows:
\begin{gather*}
logit(\gamma_{ij}) =\alpha_j+\beta z_i + \theta_{z_i} x_i
\end{gather*}
where $logit(\gamma_{ij} = \log(\gamma_{ij}/(1-\gamma_{ij}))$ are the log odds of the probability that $Y_i$ is less than $j$. The parameters $\alpha_j$ determine the cumulative probabilities of outcomes for the reference group (the control arm in the current study). The variable $z_i$ indicates whether the subject is enrolled in the current trial $z_i = 0$ or is part of the external cohort $z_i = 1$ and the parameter $\beta$ captures the differences in the distribution of the outcome for the control group from the external data. The variable $x_i$ indicates treatment assignment where $x_i = 0$ is for control and $x_i = 1$ for treatment and the parameter $\theta_z$ is the log odds ratio comparing treatment to control for population $z$ where $z = 0$ indicates current trial and $z = 1$ indicates the external cohort. 

The priors on the treatment effects $\theta_z$ are as follows:
\begin{gather*}
\theta_{z} \sim N(\mu_\theta, \tau^2_\theta), \, z=0, 1\\
\mu_\theta \sim N(0, 1)\\
\tau^2_\theta \sim InvGamma(0.125, 0.00281)
\end{gather*}

They are nested in a hierarchical prior with a shared mean and shared variance. The prior on the variance of the treatment effects, $\tau^2_\theta$ is set to have a weight of 0.25 and a central value of 0.15. This indicates that we believe the two treatment effects are likely to be similar (variance on average of 0.15) but with less weight so that the data can dictate the degree of borrowing. 

\section{Longitudinal Model}
The longitudinal modeling example uses a joint analysis of 90-day and 30-day outcomes. Using similar notation to the borrowing example, let $Y_{i,90} \in \{0, 1, 2, 3, 4, 5\}$ and $Y_{i,30} \in \{0, 1,2,3,4,5\}$  denote the 90-day and 30-day outcomes for patient $i = 1, ...., N$. The 90-day outcomes are modeled using a Bayesian ordinal regression
\begin{gather*}
logit(\gamma_{ij}) =\alpha_j+\theta x_i 
\end{gather*}
where $\gamma_{ij} = Pr(Y_{i, 90} \le j)$, $x_i \in {0, 1}$ indicates control or treatment, and the parameter $\theta$ is the treatment effect (log odds ratio). The 30-day outcomes are modeled conditionally depending on the 90-day outcomes.
\begin{gather*}
\Pr\left(Y_{i,30} = i |Y_{i,90} = j\right) =  \rho_{i,j}
\end{gather*}
The parameter $\rho_{i,j}$ is the reverse transition probability of observing outcome $i$ at 30-days given a 90-day outcome of $j$. A similar approach was used Influence of Cooling duration on Efficacy in Cardiac Arrest Patients (ICECAP) study (\href{https://clinicaltrials.gov/study/NCT04217551}{NCT04217551}).  Uninformative Dirichlet priors are placed on the transition probabilities $\rho_{j} = \{\rho_{1,j}, \ldots, \rho_{6,j}\}$.
\begin{gather*}
\rho_{j} \sim  \text{Dirichlet}\left(1/6, ..., 1/6\right) 
\end{gather*}
Together, these elements form a joint distribution of $(Y_{i,30}, Y_{i,90})$ conditional on the parameters $\alpha, \rho$ and $\theta$ which can be sampled from using Markov chain Monte Carlo.  For patients that have not completed the 90-day visit, the data are imputed from the conditional distribution $Y_{i,90} | \left(Y_{i,30}, \alpha, \rho, \theta\right)$ at each step of the Markov chain.
\section{Additional Results for the Dichotomous Example}

As shown in the main paper, the approximate predictive probability works very well in the dichotomous endpoint case, showing almost identical values between the aPP and iPP. 
There is one important caveat to this case: when the normal approximation to the binomial approximation does not work well. 
When the rate of an outcome is very low, the normal approximation used to calculate the p-value is less accurate. Because the aPP calculation relies on this p-value, it does not work as well in these cases. 

To illustrate this, we consider a wider range of possible annual rates: 0.05, 0.10, and 0.50 per year.
These are considered the base rates without treatment in three control scenarios.
For each control scenario, we simulate one of two treatment effects: no treatment effect (zero difference) and a moderate treatment effect (0.025, -0.05, or -0.15, respectively). 
For each combination, we generate 10,000 simulated trials using FACTS Version 6.5 (Berry Consultants, 2023) and process using R Statistical Software (v4.3.0; \cite{R}). Figures are made in R with ggplot2 \cite{Wickham2016}.

\begin{figure}[!htbp]
\caption{Recruitment (Sample Size) Decisions. Bayesian imputed predictive probability and approximate predictive probability across event probabilities in 10,000 simulated trials. Across event rates, the decisions made by the iPP and aPP align in the vast majority of cases. 
When there are very low event rates (2.5\%) and the normal approximation to the binomial doesn't work well, the aPP overestimates the predictive probability. However, the accuracy of the approximation is still good near 0.90 and 0.05, where decisions are typically made.}
\centering
\label{fig:dichotomous}
\includegraphics[width=\textwidth]{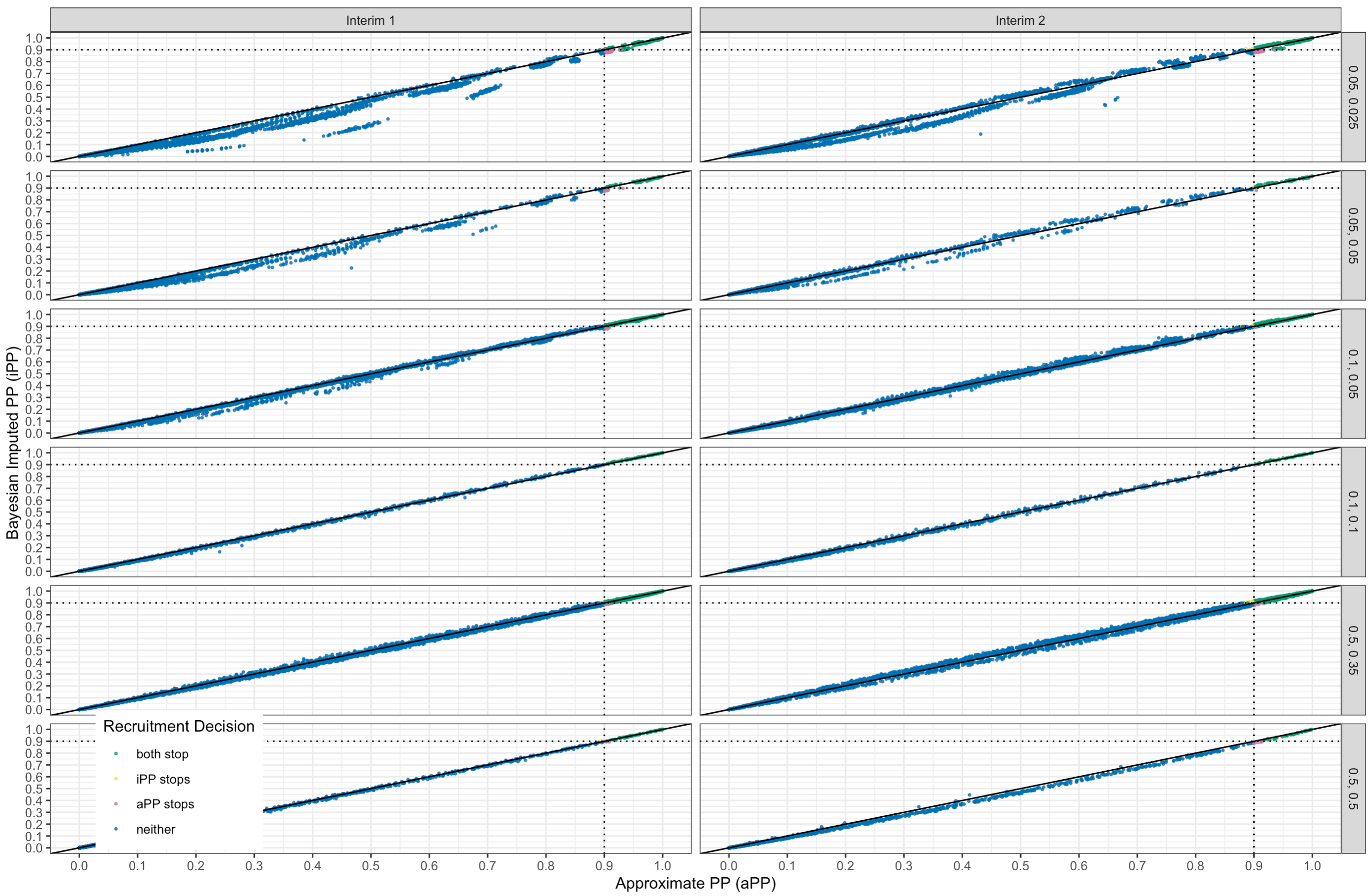}
\end{figure}

Results are displayed in Table \ref{tab:dichotomous} and Figure \ref{fig:dichotomous}
Table~\ref{tab:dichotomous} shows that the two predictive probability approaches make the same decisions about sample size and futility stopping in the vast majority of cases (97.68\%-99.94\%).
These similarities can be seen visually in Figure \ref{fig:dichotomous}, where green and blue dots indicate agreement, and the few pink and yellow dots indicate trials where the two approaches led to different interim decisions.

\begin{figure}[!htbp]
\caption{Bayesian imputed predictive probability and approximate predictive probability across event probabilities in 10,000 simulated trials. Across event rates, the decisions made by the iPP and aPP align in the vast majority of cases. 
When there are very low event rates (2.5\%) and the normal approximation to the binomial doesn't work well, the aPP overestimates the predictive probability. However, the accuracy of the approximation is still good at the tails (near 0.90 and 0.05), where decisions are typically made.}
\centering
\label{fig:small_counts}
\includegraphics[width=\textwidth]{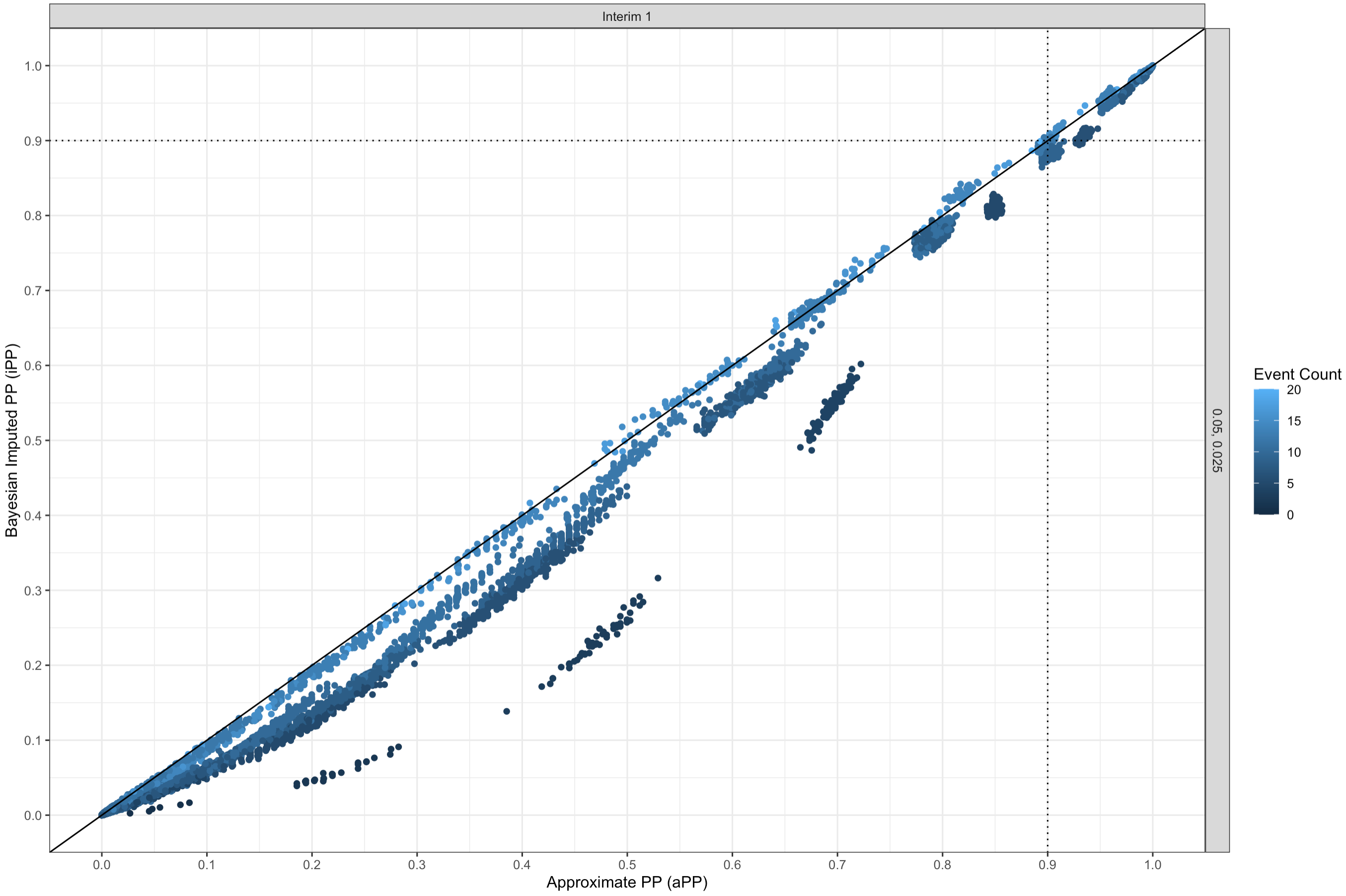}
\end{figure}

\begin{figure}[!htbp]
\caption{Futility Decisions. Comparing Bayesian imputed predictive probability with approximate predictive probability for futility decisions across event probabilities in 10,000 simulated trials. Vertical labels on right wide indicate the rates for each scenario, with the control group rate first and treatment group rate second.}
\centering
\label{fig:dichotomous_fut}
\includegraphics[width=\textwidth]{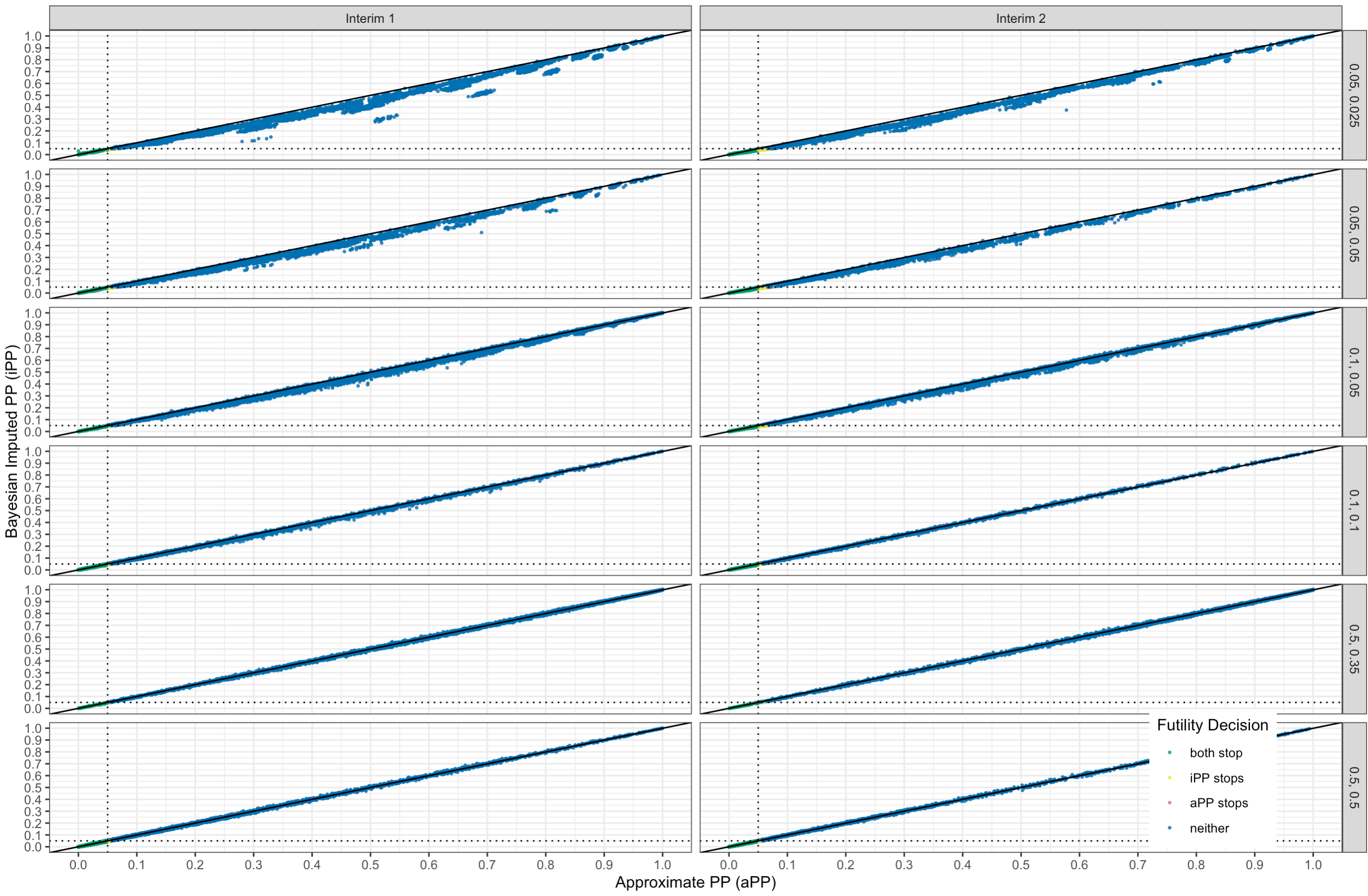}
\end{figure}

When the event rates are very low (control $p$ = 0.05, treatment $p$ = 0.025 in the top left panel in Figures ~\ref{fig:dichotomous} and~\ref{fig:dichotomous_fut}), the approximation did not work as well. Here, event counts were generally below 10 total (combining control and treatment) at the first and second interim (Figure~\ref{fig:small_counts}). 

In summary, when event counts at an interim are very low and the normal approximation to the binomial does not work well, the aPP will also not work well, since it relies on the resulting p-value and we wouldn't recommend its use.

\begin{table}

\caption{Dichotomous Primary Endpoint: Interim Decisions}
\centering
\small
\begin{tabular}[t]{r|r|r|r|r|r|r|r|r}
\hline
\multicolumn{2}{c|}{Rate} & \multicolumn{1}{c|}{Interim} & \multicolumn{3}{c|}{Success Decisions} & \multicolumn{3}{c}{Futility Decisions} \\
\cline{1-2} \cline{3-3} \cline{4-6} \cline{7-9}
Control & Treatment &   & iPP stop & aPP stop & agree & iPP stop & aPP stop & agree\\
\hline
0.05 & 0.025 & Interim 1 & 0.0623 & 0.0729 & 0.9962 & 0.1947 & 0.1909 & 0.9962\\
\hline
0.05 & 0.025 & Interim 2 & 0.0886 & 0.0994 & 0.9768 & 0.2040 & 0.1808 & 0.9768\\
\hline
0.05 & 0.050 & Interim 1 & 0.0084 & 0.0098 & 0.9915 & 0.5546 & 0.5461 & 0.9915\\
\hline
0.05 & 0.050 & Interim 2 & 0.0185 & 0.0181 & 0.9689 & 0.5088 & 0.4777 & 0.9689\\
\hline
0.10 & 0.050 & Interim 1 & 0.1784 & 0.1794 & 0.9954 & 0.0971 & 0.0925 & 0.9954\\
\hline
0.10 & 0.050 & Interim 2 & 0.1760 & 0.1767 & 0.9943 & 0.1046 & 0.0990 & 0.9943\\
\hline
0.10 & 0.100 & Interim 1 & 0.0080 & 0.0079 & 0.9878 & 0.5595 & 0.5491 & 0.9878\\
\hline
0.10 & 0.100 & Interim 2 & 0.0141 & 0.0143 & 0.9907 & 0.4909 & 0.4825 & 0.9907\\
\hline
0.50 & 0.350 & Interim 1 & 0.4760 & 0.4819 & 0.9994 & 0.0138 & 0.0134 & 0.9994\\
\hline
0.50 & 0.350 & Interim 2 & 0.4336 & 0.4287 & 0.9988 & 0.0241 & 0.0233 & 0.9988\\
\hline
0.50 & 0.500 & Interim 1 & 0.0083 & 0.0087 & 0.9934 & 0.5772 & 0.5724 & 0.9934\\
\hline
0.50 & 0.500 & Interim 2 & 0.0111 & 0.0123 & 0.9930 & 0.4970 & 0.4929 & 0.9930\\
\hline
\end{tabular}
\label{tab:dichotomous}
\end{table}

\clearpage

\bibliography{mybib.bib}

\begin{thebibliography}{10}
\providecommand \doibase [0]{http://dx.doi.org/}%

\bibitem{berry1989monitoring}
Berry DA. Monitoring accumulating data in a clinical trial. {\it Biometrics} 1989\string: 1197--1211.

\bibitem{Broglio2014}
Broglio KR, Connor JT, Berry SM. Not Too Big, Not Too Small: A Goldilocks Approach To Sample Size Selection. {\it Journal of Biopharmaceutical Statistics} 2014\string; 24(3)\string: 685-705.

\bibitem{Saville2014}
Saville BR, Connor JT, Ayers GD, Alvarez J. The utility of Bayesian predictive probabilities for interim monitoring of clinical trials. {\it Clinical Trials} 2014\string; 11(4)\string: 485-493.

\bibitem{Li2023}
Li W, Cornelius V, Finfer S, Venkatesh B, Billot L. Adaptive designs in critical care trials: a simulation study. {\it BMC Med Res Methodology} 2023\string; 23(1)\string: 236.

\bibitem{Mehta2011}
Mehta CR, Pocock SJ. Adaptive increase in sample size when interim results are promising: A practical guide with examples. {\it Statistics in Medicine} 2011\string; 30(28)\string: 3267-3284.

\bibitem{Pocock1977}
Pocock SJ. {Group sequential methods in the design and analysis of clinical trials}. {\it Biometrika} 1977\string; 64(2)\string: 191-199.

\bibitem{Jennison2000}
Jennison C, Turnbull BW. {\it Group sequential methods with applications to clinical trials}.
\newblock Chapman \& Hall/CRC .
\newblock 2000.

\bibitem{Whitehead1997}
Whitehead J. {\it The design and analysis of sequential clinical trials}.
\newblock Wiley .
\newblock 1997.

\bibitem{Vaart1998}
Vaart AWvd. {\it Asymptotic Statistics}.
\newblock Cambridge Series in Statistical and Probabilistic MathematicsCambridge University Press .
\newblock 1998

\bibitem{Satoshi2008}
Morita S, Thall PF, Müller P. Determining the Effective Sample Size of a Parametric Prior. {\it Biometrics} 2008\string; 64(2)\string: 595-602.
\newblock \href {\doibase https://doi.org/10.1111/j.1541-0420.2007.00888.x} {doi: https://doi.org/10.1111/j.1541-0420.2007.00888.x}

\bibitem{Neuenschwander2010}
Neuenschwander B, Capkun-Niggli G, Branson M, Spiegelhalter DJ. Summarizing historical information on controls in clinical trials. {\it Clinical Trials} 2010\string; 7(1)\string: 5-18.

\bibitem{Viele2014}
Viele K, Berry S, Neuenschwander B, et al. Use of historical control data for assessing treatment effects in clinical trials. {\it Pharmaceutical Statistics} 2014\string; 13(1)\string: 41-54.
\newblock \href {\doibase https://doi.org/10.1002/pst.1589} {doi: https://doi.org/10.1002/pst.1589}

\bibitem{SHINE}
Bruno A, Durkalski VL, Hall CE, et al. The Stroke Hyperglycemia Insulin Network Effort (SHINE) Trial Protocol: A Randomized, Blinded, Efficacy Trial of Standard vs. Intensive Hyperglycemia Management in Acute Stroke. {\it International Journal of Stroke} 2014\string; 9(2)\string: 246-251.
\newblock \href {\doibase 10.1111/ijs.12045} {doi: 10.1111/ijs.12045}

\bibitem{Broglio2022}
Broglio K, Meurer WJ, Durkalski V, et al. {Comparison of Bayesian vs Frequentist Adaptive Trial Design in the Stroke Hyperglycemia Insulin Network Effort Trial}. {\it JAMA Network Open} 2022\string; 5(5)\string: e2211616-e2211616.

\bibitem{Finkelstein1999}
Finkelstein DM, Schoenfeld DA. Combining mortality and longitudinal measures in clinical trials. {\it Statistics in Medicine} 1999\string; 18(11)\string: 1341-1354.

\bibitem{Berry2013}
Berry JD, Miller R, Moore DH, et al. The Combined Assessment of Function and Survival (CAFS): A new endpoint for ALS clinical trials. {\it Amyotrophic Lateral Sclerosis and Frontotemporal Degeneration} 2013\string; 14(3)\string: 162-168.

\bibitem{Ferreira2020}
Ferreira JP, Jhund PS, Duarte K, et al. Use of the Win Ratio in Cardiovascular Trials. {\it JACC: Heart Failure} 2020\string; 8(6)\string: 441-450.

\bibitem{R}
{R Core Team} . {\it R: A Language and Environment for Statistical Computing}. R Foundation for Statistical Computing; Vienna, Austria:  2021.

\bibitem{Wickham2016}
Wickham H. {\it ggplot2: Elegant Graphics for Data Analysis}.
\newblock Springer-Verlag New York .
\newblock 2016.

\end{thebibliography}
\end{document}